\documentclass{article}
\usepackage{PRIMEarxiv}

\usepackage[utf8]{inputenc} 
\usepackage[T1]{fontenc}    
\usepackage{url}            
\usepackage{booktabs}       
\usepackage{amsfonts}       
\usepackage{nicefrac}       
\usepackage{microtype}      
\usepackage{lipsum}
\usepackage{fancyhdr}       
\usepackage{graphicx}       
\graphicspath{{media/}}     
\usepackage{listings}
\usepackage[whole]{bxcjkjatype} 
\usepackage[unicode]{hyperref} 
\title{Embedding Method for Knowledge Graph with Densely Defined Ontology}

\usepackage{graphicx}
\usepackage{listings}
\usepackage{hyperref}

\author{Takanori Ugai\\
{\small Fujitsu Limited.}\\
{\small 4-1-1 Kamikotanaka Nakaharaku Kawasaki Kanagawa, 211-8588, Japan}\\
{\tt\small ugai@fujitsu.com}
}

\begin{document}
\maketitle

\begin{abstract}
Knowledge graph embedding (KGE) is a technique that enhances knowledge graphs by addressing incompleteness and improving knowledge retrieval. A limitation of the existing KGE models is their underutilization of ontologies, specifically the relationships between properties. This study proposes a KGE model, TransU, designed for knowledge graphs with well-defined ontologies that incorporate relationships between properties.  The model treats properties as a subset of entities, enabling a unified representation. We present experimental results using a standard dataset and a practical dataset.
\end{abstract}


\section{Introduction}
Knowledge graphs, which are often represented using the Resource Description Framework (RDF), are assumed to express true statements.  However, many real-world knowledge graphs such as DBpedia\cite{10.1007/978-3-540-76298-0_52} are incomplete.  For example, a significant portion of the entities in DBpedia lack proper assignments. For instance, 27\% of entities classified as universities are not classified as organizations, which is a more general type they should belong to\cite{caminhas2019dbpedia}  To address this incompleteness and derive new knowledge, techniques are used to compensate for missing information, such as applying inference rules or integrating them with other knowledge graphs and databases\cite{paulheim2017knowledge,10.5555/2401764}. 
Knowledge graph embedding (KGE) is an example of such a technique.  KGE represents the relationships between subjects, properties, and objects in a knowledge graph as vectors, enabling approximation through inner product similarity\cite{DBLP:journals/corr/Nguyen17a}.  KGE is expected to aid in predicting missing information in knowledge graphs, and has been the subject of much research.  However, achieving the high accuracy required for automated data addition (e.g., 90\% in Wikidata) remains a challenge \cite{DBLP:journals/corr/Nguyen17a,paulheim2017knowledge}.

A limitation of the existing KGE techniques is their underutilization of ontologies.  Ontologies provide the semantic context of knowledge graphs written in RDF; however, current methods primarily focus on vector representations of node-property-node triplets using formal neural techniques.  To address this, this study developed a KGE technique that incorporates the relationships between ontology properties. 

Traditional KGE methods treat knowledge graphs as sets of nodes, properties, and definitions.  However, RDF allows properties to have their own definitions, effectively allowing them to act as nodes.  This proposal treats properties as a part of a node, creating a model that accounts for statements about properties.  This approach enables a more unified representation, where nodes connected by semantically similar properties (e.g., translations) are represented closely in the embedding space

Section 2 illustrates the limitations of existing KGE methods using TransE as an example.  Section 3 details the proposed knowledge graph embedding method.  Section 4 compares the experimental results of TransU with those of existing KGE methods.  Finally, Section 5 discusses the results, summarizes the study, and outlines future work.

\section{Problems with existing methods}

TransE\cite{NIPS2013_5071}, a typical knowledge graph embedding method, learns that if the subject, property, and object representation vectors of a knowledge graph are $v_s$, $v_p$, and $v_o$, respectively, then in a positive example triplet, the relationship $v_s + v_p = v_o $ holds.
Various other methods exist, such as models that restrict space according to the relationship (TransH)\cite{wang2014knowledge}, models based on matrix transformations (TransR) \cite{10.5555/2886521.2886624}, and models that convert to a complex space (ComplEx) \cite{trouillon2016complex}.

Knowledge graph embedding models aim to represent entities and relations in a knowledge graph as low-dimensional vectors. This facilitates the application of machine learning algorithms to graph-structured data for tasks, such as link prediction, where the goal is to infer missing relationships between entities. These models are crucial for dealing with the incompleteness of knowledge graphs, which are often represented using a Resource Description Framework (RDF).

Several knowledge graph embedding techniques have been developed, each with its own approach.

\begin{itemize}
    \item \textbf{TransE (Translating Embeddings):} As mentioned in the original text, TransE models relationships as translations in the embedding space. Given a triple (subject, relation, object), TransE aims to learn embeddings such that the embedding of the subject plus the embedding of the relation is close to the embedding of the object ($v_s + v_p \approx v_o$). \cite{NIPS2013_5071}

    \item \textbf{TransH (Translation with Hyperplanes):} TransH introduces relation-specific hyperplanes. Instead of a single vector for each relation, a relation vector and a hyperplane are used for each relation. This allows entities to have different representations for different relations. \cite{wang2014knowledge}

    \item \textbf{TransR (Translation in Relation-Specific Spaces):} TransR extends TransH by projecting entities into relation-specific vector spaces before applying the translation. This allows the model to better capture the diverse roles of entities in different relations. \cite{10.5555/2886521.2886624}

    \item \textbf{ComplEx (Complex Embeddings):} ComplEx represents entities and relations as complex-valued vectors. It uses complex-valued vector operations to model both symmetric and asymmetric relations. \cite{trouillon2016complex}

    \item \textbf{RDF2Vec:} This approach generates vector representations for entities in RDF graphs. It uses graph-based random walks to capture the structural context of entities, and then applies techniques such as skip-gram to learn embeddings. RDF2Vec\cite{Ristoski2019} supports knowledge graph updates.

    \item Other notable models include DistMult\cite{Yang2014}, HolE\cite{Nickel2016}, and those based on neural networks.
\end{itemize}

In these models, a graph is defined as a set of pairs of elements of the set of nodes and branches in the form $G = (E, V)$.
It also indicates that E and V are independent sets.
However, the RDF states that a branch can be a node.

\begin{center}
\begin{tabular}{ccc}
    ex:A & exp:p1 & ex:C . \\
    exp:p1 & a & exp:T1 .
\end{tabular}
\end{center}

Properties in the RDF have a structure defined in the form of an ontology.
For example, the description above defines a type of property $exp:p1$.
The existing model defines the representation vectors for $exp:p1$ in the first and second lines, respectively, as different vectors.

Consequently, existing embeddings fail to predict links using an ontology structure.
For example, it becomes difficult to infer links when the type is not defined in $exp:p1$.

\begin{figure}
  \centering
\includegraphics[width=0.3\linewidth]{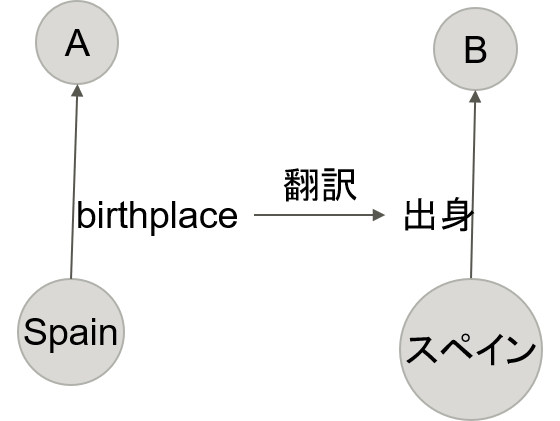}
  \caption{bad example of learning}\label{Fig1}
\end{figure}

Fig. \ref{Fig1} is an example where existing knowledge graph embedding methods do not learn well.
The RDF form is as follows:

\begin{center}
\begin{tabular}{ccc}
      A & birthplace & Spain . \\
      B & 出身  & スペイン . \\
      birthplace & 翻訳 & 出身 .
\end{tabular}
\end{center}

In this RDF, we intend to predict whether A's birthplace is Spain and B's birthplace is Spain, because 出身地 and birthplace are related by translation.

\begin{figure}
  \centering
\includegraphics[width=0.3\linewidth]{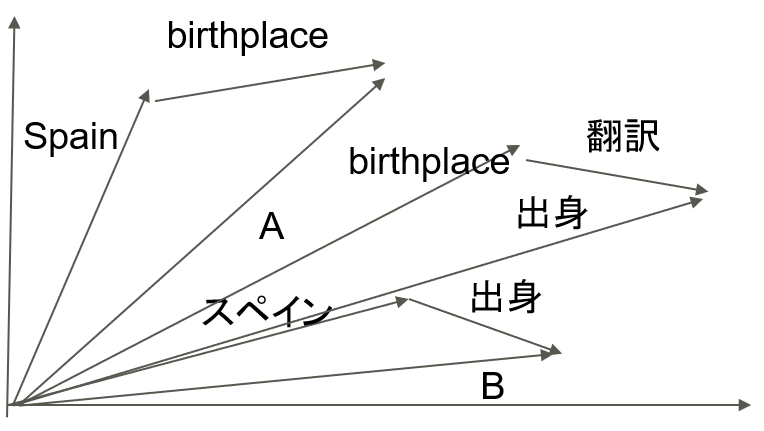}
  \caption{Example expressed as a two-dimensional vector}\label{Fig2}
\end{figure}

For simplicity, Fig. \ref{Fig2} shows a two-dimensional representation of the RDF of Fig. \ref{Fig1}, using TransE.
Because the birthplace and origin of the property and the birthplace and origin of the entity are represented by different vectors 
A, and Spain are not related. Therefore, it is difficult to predict whether A's birthplace is Spain.

\begin{figure}
  \centering
\includegraphics[width=0.3\linewidth]{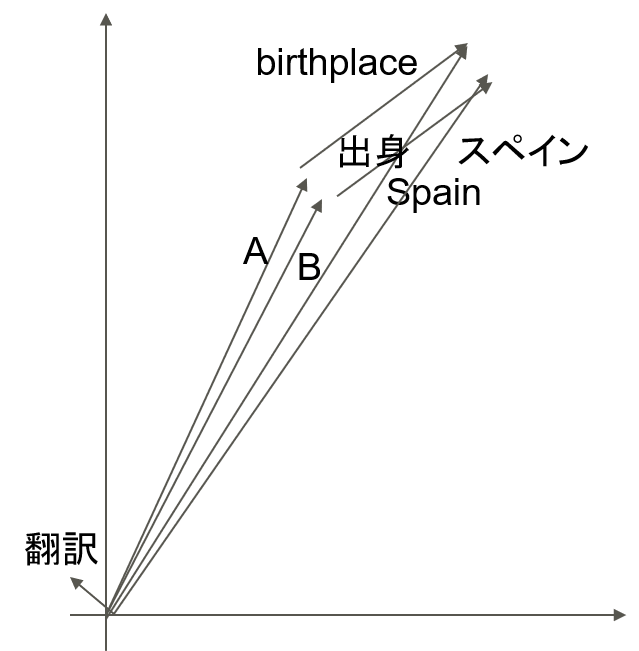}
  \caption{Expected vector representation}\label{Fig3}
\end{figure}

To be able to predict that A's and B's birthplace are Spain, we need that the vectors of birthplace and 出身 are close,
as shown in Fig. \ref{Fig3}.

In Fig. \ref{Fig3}, A and B have the same birthplace, so they are represented by close vectors, and
Spain and Spain are represented by close vectors owing to their translation relations.
Consequently, A and Spain were represented by close vectors. Using vector distances,
We can deduce that A was from Spain.

\section{Proposal: TransU}

This study introduced a knowledge graph embedding method called TransU.
In TransU, the subjects, properties, and objects within the knowledge graph are unified into a single-entity set, denoted as E. To formalize this, we define two subsets: $E1$, representing the set of subjects and objects, and $E2$, representing the set of properties. Then, the graph structure is defined as $G = (E1, E2)$.

The key idea is that properties are treated as a subset of entities ($E2 \subset E1 \subset E$).
Whereas the learning algorithm for the representation vectors can be any existing knowledge graph embedding method, TransU imposes a constraint: properties, when acting as entities, must be represented by the same vector.

\begin{figure}
  \centering
\includegraphics[width=0.6\linewidth]{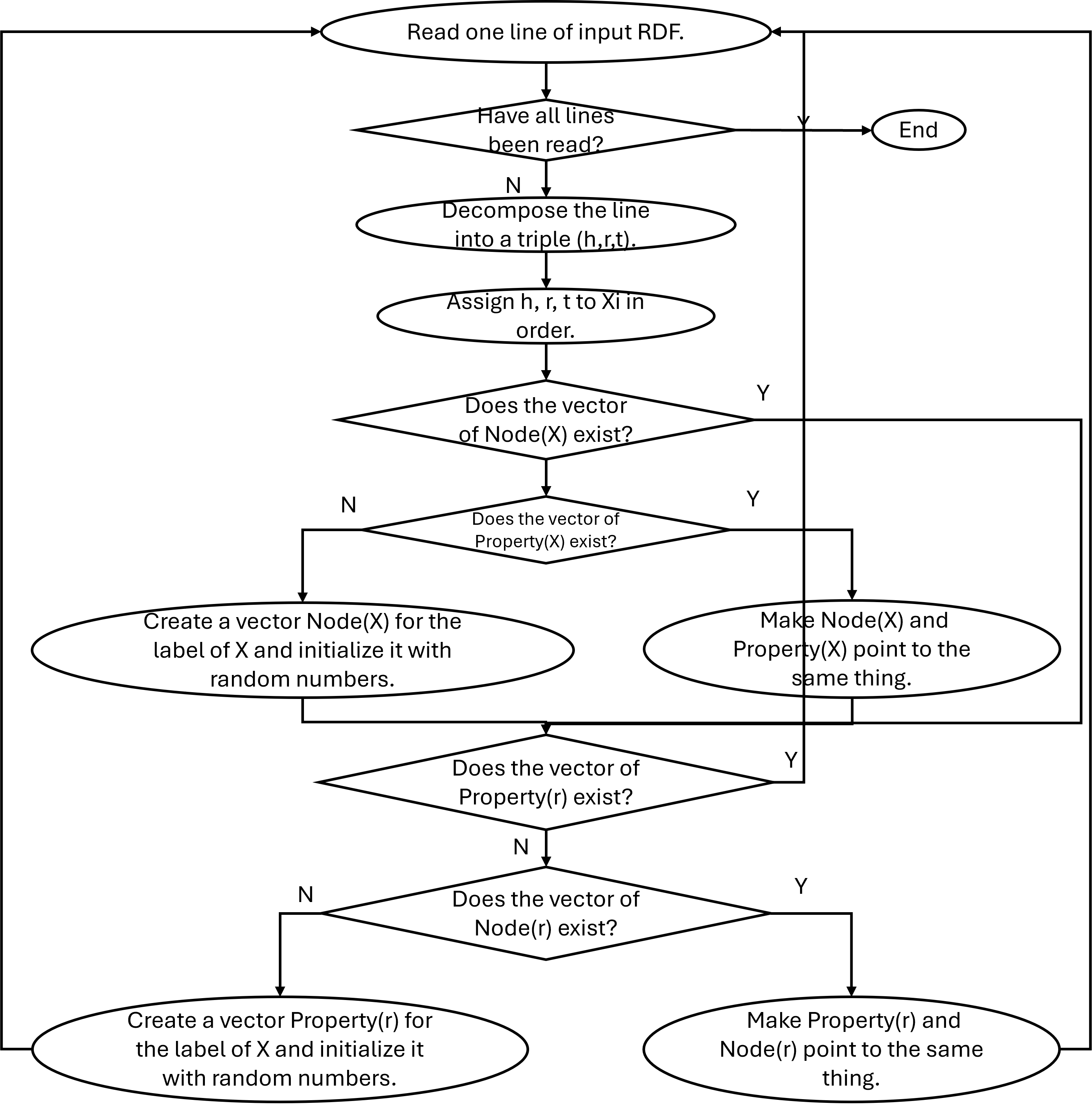}
  \caption{Initialisation algorithm for vector representation}\label{Fig4}
\end{figure}

To ensure compatibility with the existing knowledge graph embedding methods, the dimensions of the entity and property vectors must be consistent.
When integrating TransU with existing methods, the algorithm illustrated in Figure \ref{Fig4} is applied specifically during the initialization phase of the representation vector learning.

Essentially, when a property is used as an entity, the same vector is used for representation.

\section{Experiment}
To evaluate the proposed TransU method, we conducted experiments using two datasets: FB15K and a knowledge graph derived from a mystery novel, referred to as "speckled string."    

\begin{description}
\item[FB15K\cite{toutanova-chen-2015-observed}]: This dataset, extracted from Freebase, is commonly used for evaluating knowledge graph embedding methods.  It contained 592,213 triples, 14,951 entities, and 1,345 properties.  Importantly, FB15K does not include relationships between the properties.  In our experiments, TransU was evaluated in combination with TransE and TransH on the dataset.    
\item[Speckled String\cite{kawamura2019report}]: This dataset is derived from a mystery novel and is considered to have a rich definition of properties.  It contains 4,342 triples, 2,234 entities, and 43 properties, with 436 triples, where a property is a subject or object.  The entire dataset was used for both training and testing.    
\end{description}

\subsection*{Experimental Setup}

For both datasets, the entities and properties were learned as vectors in the same space during the training.  However, during the evaluation, entities and properties were distinguished, and properties that were not entities were excluded when calculating the rank of correct answers.  For the "speckled string" dataset, properties were treated as entities, but entities were not treated as properties during the evaluation.  The size of the representation vector was set to 200 for both the datasets.  For ComplEx, the vector size is 100 (real and imaginary components).  The learning rates were set to 0.001 for TransE and TransH and 0.01 for ComplEx, using the Adam optimizer.  All the algorithms were trained for 1000 cycles.

\begin{table*}
  \centering
\caption{Results of evaluation experiment using FB15K}\label{Table1}
 \begin{tabular}{ccccc}
Model	&MeanRank(Raw)	&MeanRank(Filter)	& Hit@10(Raw)	& Hit@10(Filter) \\ \hline
TransE(paper)	& 243	& 125	& 34.9	& 47.1 \\ \hline
TransH(paper)	& 212	& 87	& 45.7	& 64.4 \\ \hline
TransE(Base)    & 244   & 132  & 32.6 & 46.1 \\ \hline
TransH(Base)    & 202   & 84   & 42.3 & 54.4 \\ \hline
TransU(TransE):Avg  & 260 & 30.3 & 42.3 & 52.0 \\ \hline
TransU(TransE):Best & 202 & 28.2 & 40.0 & 42.4 \\ \hline
TransU(TransH):Avg  &220 &48.3 &66.6 & 70.8 \\ \hline
TransU(TransH):Best &198 &40.4 &38.0 & 40.2 
\end{tabular}
\end{table*}

\subsection*{Results}

\begin{itemize}
\item FB15K: Table \ref{Table1} shows the results of the evaluation experiments.  The table presents the performances of TransE and TransH, both in their original form and in combination with TransU.  The results are shown as the average score of ten trials and the best score of ten trials.  The average score of the ten trials was slightly lower than that of the baseline methods (TransE and TransH).    
\item Speckled String: Table \ref{Table2} presents the results of the evaluation experiments.  The table shows the performances of TransE, TransH, ComplEx, and their combinations with TransU.  The combination of TransU and ComplEx yielded the best results. 
\end{itemize}

\begin{table}
  \centering
\caption{Results of evaluation experiments using ``speckled string''}\label{Table2}
 \begin{tabular}{ccc}
Model	&MeanRank	& Hit@10 \\ \hline
TransE    & 2.10   & 89  \\ \hline
TransH    & 2.02   & 84   \\ \hline
ComplEx    & 1.47   & 92 \\ \hline
TransU(TransE)  & 2.00 & 74 \\ \hline
TransU(TransH) & 1.98 & 82 \\ \hline
TransU(CompEx) & 1.42 & 92
\end{tabular}
\end{table}

\section{Discussion}
The experiments with FB15K showed a slightly lower average score for TransUthano than for the baseline methods.  This is likely because FB15K lacks a rich ontology of properties, meaning that it does not have many properties that also function as entities.  Consequently, the benefits of TransU, which is designed to leverage such property relationships, were not fully realized in this dataset.    

Furthermore, the evaluation methodology involved distinguishing between entities and properties and excluding properties that were not entities when calculating the ranking of correct answers.  During training, however, entities and properties were not distinguished, leading to all combinations being considered correct answers.  This could have introduced noise, as some entity-property pairs might have been coincidentally close in the vector space, but not genuinely related.  These potentially noisy combinations were excluded from evaluation.    

In contrast, the experiments with the "speckled string" data treated properties as entities during evaluation, but not versaa.  This led to the inclusion of less natural RDF triads, such as entity-type-literal and aspotentiallyl-correct answers.  The accuracy of the model could potentially be improved by incorporating literals as subjects or by using ontology to define the range or domain types for properties.

\section{Summary and Future works}
This study proposes TransU, a knowledge graph embedding method designed to leverage the relationships between ontology properties, addressing a key limitation of existing KGE models.  The proposed model unifies the representation of subjects, properties, and objects by treating properties as a subset of entities, thereby enabling closer relationships between nodes connected by semantically similar properties.  Evaluation experiments demonstrated that TransU, when combined with existing embedding methods, can improve the accuracy of knowledge graphs with richer property definitions.  However, the improvement was marginal for knowledge graphs with fewer interproperty relationships, highlighting the need for enhanced adaptability across diverse datasets.    

Future studies should focus on several key areas to enhance the robustness and applicability of TransU.

\begin{itemize}
\item {\bf Improving Generalizability:} Develop strategies to extend TransU's effectiveness to knowledge graphs with varying degrees of ontological richness. This could involve hybrid approaches that dynamically combine TransU with other KGE methods or techniques to infer property relationships from the data.
\item {\bf Refining Evaluation Methodologies:} Implement more rigorous evaluation protocols that maintain a clear distinction between entities and properties during both training and evaluation. This includes exploring new evaluation metrics that specifically assess the prediction of property-property relationships and using separate datasets for training and evaluation.
\item {\bf Modeling Nuanced Property Relationships:} Investigate techniques to capture the subtleties of property relationships beyond simple similarity. For example, explore the use of attention mechanisms to weigh the importance of different relationships or employ separate embedding spaces to represent distinct relationship types (e.g., hierarchical, associative, and antonymous).
\item {\bf Enhancing Data Type Handling:} Expand TransU to effectively handle diverse data types, including literals, dates, and numerical values. This could involve type-specific embedding layers or transformations to ensure the accurate representation and processing of different data modalities.
\item {\bf Scalability and Complexity Analysis:} Conduct a thorough analysis of TransU's computational complexity and scalability, comparing it with existing KGE methods. This analysis should identify potential bottlenecks and guide the development of optimizations to handle large-scale knowledge graphs.
\end{itemize}

\bibliographystyle{splncs04}

\bibliography{lod}

\end{document}